\documentclass[useAMS,usenatbib,usegraphicx,referee]{mn2e}
\usepackage{graphicx,amssymb,amsmath,times,verbatim}
\usepackage{color}
\usepackage{subfigure}
\usepackage{upgreek}

\usepackage{adjustbox}
\usepackage{lipsum}

\title[Temporal and Spectral Study of 3C\,279]{Study on Temporal and Spectral behavior of 3C\,279 during 2018 January flare}

\author[Z. Shah et al.]{Zahir Shah$^{1, 2}$\thanks{Email: shahzahir4@gmail.com}, V. Jithesh$^{1}$, S. Sahayanathan$^{3}$, Ranjeev Misra$^{1}$
and Naseer Iqbal$^{2}$ \\
$^{1}$ Inter-University Center for Astronomy and Astrophysics, PB No.4, Ganeshkhind, Pune-411007, India\\
$^{2}$ Department Of Physics, University of Kashmir, Srinagar-190006, India \\
$^{3}$ Astrophysical Sciences Division, Bhabha Atomic Research Centre, Mumbai-400085, India}

\begin{document}
\date{}
\pagerange{\pageref{firstpage}--\pageref{lastpage}} \pubyear{2016}
\maketitle
\label{firstpage}
\begin{abstract}
We present a detailed temporal and spectral study of the blazar 3C\,279 using multi-wavelength observations from \emph{Swift}-XRT, \emph{Swift}-UVOT and \emph{Fermi}-LAT during a flare in 2018 January. The temporal analysis of $\gamma$-ray light curve indicates a lag of $\sim 1$\,d between the 0.1--3\,GeV and 3--500\,GeV  emission. Additionally, the $\gamma$-ray light curve shows asymmetry with slow rise--fast decay in energy band 0.1--3\,GeV and fast rise--slow decay in the 3--500\,GeV band. We interpret this asymmetry as a result of shift in the Compton spectral peak. This inference is further supported by the correlation studies between the flux and the parameters of the log-parabola fit to the source spectra in the energy range 0.1--500\,GeV. We found that the flux correlates well with the peak spectral energy and the log-parabola fit parameters show a hard index with large curvature at high flux states.  Interestingly, the hardest index with large curvature was synchronous with a very high energy flare detected by H.E.S.S. Our study of the spectral behavior of the source suggests that $\gamma$-ray emission is most likely to be associated with the Compton up-scattering of IR photons from the dusty environment. Moreover, the fit parameters indicate the increase in bulk Lorentz factor of emission region to be a dominant cause for the flux enhancement.

\end{abstract}

\section{Introduction} 
Blazars are the special class of active galactic nuclei (AGNs) with a powerful relativistic jet of plasma 
pointing along the line-of-sight 
of the observer \citep{Blandford1979}. Emission from blazars extends from radio to $\gamma$-ray energies and are known to be the 
brightest sources in the $\gamma$-ray universe. In addition to the broad emission spectra, they are also highly variable with 
flux doubling timescale ranging from minutes to days \citep{Aharonian2007,Saito2013}. These extreme properties are usually attributed to the 
relativistic motion of
the emission region moving down the jet and are often used to constrain the source energetics \citep{Dondi1995}. 
Based on the
presence/absence of line features in their optical spectrum, blazars are further subdivided into flat spectrum 
radio quasars (FSRQs)/BL\,Lacs. 

The spectral energy distribution (SED) of blazars are characterized by two prominent peaks with the low energy 
component well understood to be synchrotron emission from a non-thermal distribution of electrons. The high 
energy component is usually attributed to synchrotron self Compton (SSC) and/or the Compton scattering of an external 
photon field (EC) by the same electron distribution \citep{Marscher1985,Dermer1992}. The external photon field can be the monochromatic photons from the broad line regions or thermal infra-red (IR) photons from the dusty torus or the emission from the accretion disk \citep{Sikora1994, Dermer1993, Boettcher1997, Ghisellini2009}. The broadband SED of BL\,Lacs can be easily interpreted as synchrotron and  SSC processes \citep{Coppi1999,Finke2008,Mankuzhiyil2011}; whereas for FSRQs, the simultaneous observations in 
X-rays and $\gamma$-rays suggest a combination of SSC and EC processes
to explain their high energy emission \citep{Sahayanathan2012,Shah2017}. Besides these lepton based emission models, the 
high energy component of blazars  has been also interpreted as an outcome of hadronic cascades \citep{Mannheim1992, Bottcher2007}.

3C\,279 is a FSRQ located at a redshift $z=0.536$ \citep{Lynds1965}. It was known to be one of the 
powerful $\gamma$-ray source in the high-energy sky since the observations by 
Energetic Gamma-Ray Experiment Telescope (EGRET) on-board the Compton $\gamma$-Ray Observatory \citep[CGRO;][]{Hartman1992}. 
After the advent of \emph{Fermi} satellite, 3C\,279 was regularly monitored in 100 MeV -- 300 GeV energies 
during various flaring states and was supplemented with the simultaneous observations in X-ray and UV/Optical
frequencies.  The source went through a series of distinct flaring events from 2013 December to 2014 April, with maximum one-day averaged $\gamma$-ray flux of $\rm (6.54\pm 0.30)\times 10^{-6}\,photons\,cm^{-2}\,s^{-1}$ recorded 
on 2014 April 03 \citep{Paliya2015a}. During this period 3C\,279 has shown a very hard $\gamma$-ray index of $1.7\pm 0.1$ in one of the flaring event, which is unusual among FSRQs \citep{Hayashida2015, Paliya2016}. Moreover, an hour timescale variability  ($\rm 1.19\pm0.36\, hr$) was observed in the $\gamma$-ray emission from 3C\,279 \citep{Paliya2015a}.  In 2015 June, 3C\,279 exhibited a record breaking outburst at GeV energies and  it reached a highest daily flux level of $\rm (2.45\pm0.05)\times 10^{-5} \,photons\,cm^{-2}\,s^{-1}$ 
\citep{Paliya2015b}. In addition to prodigious enhancement in $\gamma$-ray flux, a significant flux variability at sub-orbital timescales ($\sim 5$ min) was  observed by \emph{Fermi}-LAT for the first time \citep{Ackermann2016}. 3C\,279 is also one of the primary
blazar studied using the Whole Earth Blazar Telescope (WEBT) campaign \citep{Bottcheretal2007, Larionov2008}. The 2006 WEBT campaign of 3C\,279 at optical/IR/radio bands observed an exponential decay pattern of fluxes in B, V, R and I bands on  timescale of 12.8\,d. The results suggested a possible signature of deceleration of the emitting components in  the jet \citep{Bottcher2009}. The correlation observed between  powerful $\gamma$-ray flares and the  change in optical polarization angle strongly supports the standard one zone model \citep{Abdo2010}. On contrary, the  strong Compton dominance and minute timescale $\gamma$-ray variability in the 2015 June flaring episode pose challenges to standard one zone models model, and instead alternative models like  mirror driven clumpy jet model or/and synchrotron origin from a magnetically dominated jet etc., are suggested for the GeV $\gamma$-ray emission \citep{Ackermann2016, Vittoini2017, Pittori2018}. Further, a hadronic model was also proposed to explain the complex flux variations observed across the broadband spectrum during 2015 June flare \citep{,Romoli2017}. 3C\,279 was also the first FSRQ detected at very high energy (VHE) by Major Atmospheric Gamma-ray 
telescope with Imaging Camera \citep[MAGIC;][]{Albert2008, Aleksic2014} and this discovery raised serious 
discussions regarding the opacity of our universe to VHE $\gamma$-rays \citep{Bottcher2016, Abolmasov2017}. 
Detection of this source in VHE also indicates the presence of additional emission process in the 
high energy spectra \citep{Sikora1994,Aleksic2011,Sahayanathan2012}. Despite these intense multi-wavelength campaigns and theoretical studies, there is not yet a clear
consensus about the nature and origin of the high energy emission from 3C\,279.

In the present work, we study the January 2018 flaring activity of 3C\,279, to understand its 
temporal and spectral properties. We obtained the simultaneous information of the source in $\gamma$-ray--X-ray--optical/UV
energies using \emph{Fermi}-LAT, \emph{Swift}-XRT, and \emph{Swift}-UVOT observations. The temporal behavior
of the source is examined by studying the profile of  one-day averaged $\gamma$-ray light curve at two energy 
bands namely 0.1--3 GeV and 3--500 GeV. The inferences put forth are  justified through a detailed correlation
study of the spectral fit parameters and the observed flux. To study the spectral behavior of the source
during the flare, we extracted the broadband SED for three flux states selected from the multi-wavelength light curve. The resulting SED is investigated under simple emission model involving  synchrotron, SSC and EC processes. The paper is organized as follows: In \S \ref{sec:analysis} we outline the 
\emph{Fermi} and \emph{Swift} observations and their data analysis procedures. Following this, we present the 
results of the temporal behavior of the source during the flare 
in sections \S \ref{sec:temp_analysis} and in section \S \ref{sec:spec_analysis}, we study the spectral properties of the source in three different flux states. 
Throughout the paper, we have used a cosmology with $\rm \Omega_m = 0.27$, $\Omega_\Lambda = 0.73$ and
 $H_0=71\,\rm{km\,s^{-1}\,Mpc^{-1}}$.

 \section {Observations and Data Analysis}\label{sec:analysis}
\subsection{\emph{Fermi}-LAT} 

\begin{figure}
\centering
\includegraphics[width=0.7\textwidth,angle=0]{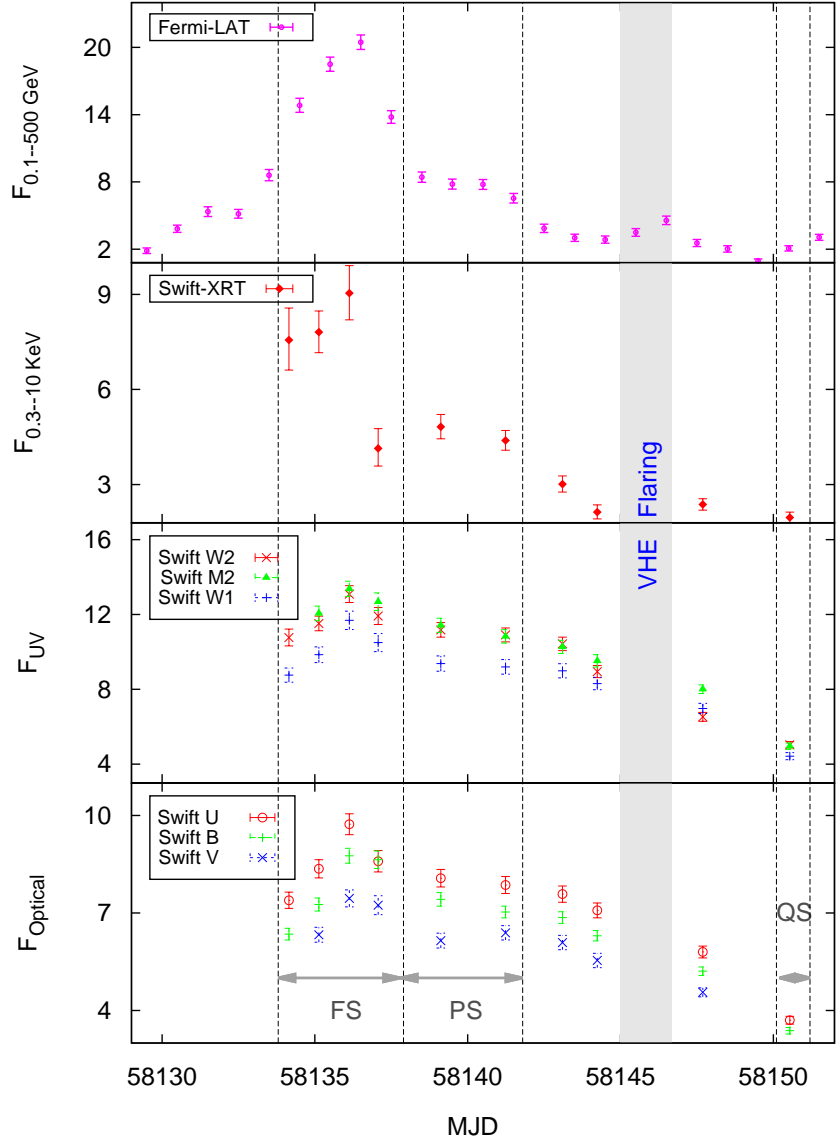}
\caption{Multi-wavelength light curve of 3C\,279 during period MJD\,58129--58152. The $\gamma$-ray flux points  are one day binned in units of $\rm 10^{-6} ph\,cm^{-2}\,s^{-1}$,  \emph{Swift}-XRT (0.3--10 keV) flux points are in units of $\rm 10^{-11}\,erg\,cm^{-2}\,s^{-1}$, and \emph{Swift}-UVOT fluxes are in units of $\rm 10^{-15}\,erg\,cm^{-2}\,s^{-1}\,A^{-1}$. The vertical dotted lines with horizontal gray arrow headed lines represents the time slots corresponding to three flux states viz. `flaring state (FS)', `plateau state (PS)' and `quiescent state (QS)' for which time averaged SEDs are obtained. The vertical solid gray region corresponds to VHE flaring time of 3C\,279  \citep{2018ATel11239....1N}.}
\label{fig:mwl}
\end{figure}

\emph{Fermi}-LAT is a pair conversion detector \citep{Atwood2009} with large effective area ($\sim 8000\,\rm cm^2/GeV$ photon) and large field-of-view ($\sim 2.4$ sr). LAT is sensitive to photons with energy ranging from 20 MeV to 500 GeV.  We collected the one month (2018 January) \emph{Fermi}-LAT data of 3C\,279  within $15^\circ$ region of interest (ROI) with center at the source position in the energy range 0.1--500 GeV. The one month  data is used to obtain the model file with significant background sources (i.e source with $TS>25$ in one month). The energy range for data collection is chosen to be $\geq 100$ MeV in order to minimize systematics. The data is analyzed with latest \emph{Fermi} {\small SCIENCE TOOLS} (v10r0p5) and using PASS8 IRFs, following standard procedures\footnote{http://fermi.gsfc.nasa.gov/ssc/data/analysis/}. The instrument response function {\small `$\rm P8R2\_SOURCE\_V6$'}, Galactic diffuse model {\small `$\rm gll\_iem\_v06.fit$'} and isotropic background model {\small `$\rm iso\_p8R2\_SOURCE\_V6\_v06.txt$'} were used to extract flux and spectra from the SOURCE class events by performing the fitting with unbinned maximum likelihood algorithm included in pylikelihood library of \emph{Fermi} {\small SCIENCE TOOLS}. The $\gamma$-ray events contaminated by the bright Earth limb were excluded using the zenith angle cut of $90^\circ$.  In the fitting procedure, the normalizations of the isotropic and Galactic diffuse emission components were kept free, whereas the index of Galactic diffuse component was fixed to its third LAT catalog (3FGL; \citealp{Acero2015}) value.  We initially carried out the likelihood analysis for full month and in the model file, we have included all the sources within $25^\circ$ ($15^\circ$ ROI and $10^\circ$ annular region) defined in the 3FGL catalog. The model parameters for the source lying within $15^\circ$ ROI were kept free, whereas the parameters of the source lying beyond $15^\circ$ were kept fixed to their 3FGL values. All the background sources with $\rm TS < 25$ were deleted from the output model file, which is finally used for the generation of light curve and the spectral analysis. Besides the proper convergence of fitting, all the flux points obtained in the light curve and spectral analysis have $TS>9$.  The $\gamma$-ray data covering the flaring period was divided into 24-hour time bin in order to obtain one-day averaged $\gamma$-ray light curve. The flux points in each time bin were obtained in the energy range of 0.1--500 GeV by fitting log parabola model to the source spectra using unbinned maximum likelihood algorithm. The obtained one-day binned $\gamma$-ray light curve is shown in the top panel of Figure \ref{fig:mwl},   which displays a well-defined peak at MJD\,58136.5 with a daily averaged flux of $\rm (2.05\pm0.06)\times 10^{-5}\,ph\,cm^{-2}\,s^{-1}$, photon index at pivot energy of $1.99\pm0.04$ and curvature parameter of $0.18\pm0.03$. 

\subsection{\emph{Swift}}

The Neil Gehrels {\it Swift} satellite \citep{2004ApJ...611.1005G} observed the flaring activity of 3C 279, happened in 2018 January, with its XRT \citep{2005SSRv..120..165B} and UVOT \citep{2005SSRv..120...95R} instruments. We selected 10 observations ( ObsID: 00035019206, 00035019210, 00035019211, 00035019213, 00035019214, 00035019218, 00035019219, 00035019220, 00035019221 and  00035019222 ) from 2018 January 17 to 2018 February 1 encompassing the flare and reprocessed the XRT data using the {\sc xrtpipeline} tool version 0.13.4, with the standard filtering and screening criteria, in the {\sc heasoft} package version 6.22.1.  The XRT data (0.3--10 keV energy band) were mostly collected in photon counting (PC) mode, except for 00035019213, 00035019214, 00035019218 and 00035019219. In such cases, we used the window timing (WT) mode data for the analysis. We used a sliding-cell detection algorithm in {\sc ximage} and detected the source in all the observations. The three observations (ObsID: 00035019206, 00035019210 and 00035019211) taken with the PC mode were affected by pile-up, where the count rate is $> 0.5 \rm~ct~s^{-1}$, while for the rest of the observations, the source count rate is low and thus no pile-up correction is required. In WT mode, the source count rate is well below the threshold level of the pile-up, thus no correction is required. 

The source and background regions were extracted from the cleaned event files with the standard grade filtering of 0--12. In the PC mode, we used a circular region with a radius of 47 arcsec; however, in case of pile-up, we prefer to use an annular region with inner radius of 6--8 arcsec and outer radius of 47 arcsec. A box region with a width of 70 arcsec and a height of 20 arcsec were used to extract the source and background in WT data. We generated the ancillary response files using the tool {\sc xrtmkarf} and used the spectral redistribution matrices available in the calibration database. The spectra obtained from individual observations were fitted with an absorbed power law (PL) model ({\tt tbabs $\times$ power law}) in {\sc xspec} \citep{1996ASPC..101...17A}, where the absorption is fixed at the Galactic value $2.05 \times 10^{20} \rm cm^{-2}$ \citep{2005A&A...440..775K}. The 0.3--10 keV unabsorbed flux (derived using the convolution model {\tt cflux}) and their uncertainties at 90 percent confidence level are shown in Figure \ref{fig:mwl}.

UVOT observations were performed with all six optical and UV filters namely, $u$, $b$, $v$, $W1$, $W2$ and $M2$. We summed the available frames of each filter with the {\sc uvotimsum} task in the HEASOFT package and obtained a single image for the corresponding filter. In some observations, a single frame is available for each filter and we used the individual frame for the analysis. We performed the source detection routine task, {\sc uvotdetect} on these summed images or individual frames using the latest UVOT CALDB version 20170922 with a threshold limit of $3\sigma$. We searched the UV-optical counterpart of 3C 279 in the UVOT images using the XRT positional uncertainty at a 90 percent confidence level, which is typically 3--4 arcsec and identified the counterpart. We extracted the source events from a circular region of 5 arcsec radius, while for the background a nearby, source-free circular region of 10 arcsec radius was used. The magnitudes in the Vega System and the corresponding flux were estimated using the {\sc uvotsource} task. The observed optical/UV flux were corrected for Galactic extinction using E(B $-$ V) = 0.029 and $R_{V} = A_{V}$/E(B $-$ V) = 3.1 following \citet{2011ApJ...737..103S}, and they are plotted in Figure \ref{fig:mwl}. 
 The active state of 3C279 has been monitored in the optical and near-IR (NIR) bands using different ground-based telescopes \citep{Dammando2018, Marchini2018, Kaur2018}. The Rapid Eye Mounting (REM) telescope performed the optical/NIR observation on 2018 January 17 with different filters \citep{Dammando2018} and measured the magnitudes in $V$ ($\sim 14.4$), $R$ ($\sim 14.0$), $I$ ($\sim 13.4$), $J$ ($\sim 12.1$), $H$ ($\sim 11.3$) and $K$ ($\sim 10.3$) bands. The SARA-KPNO telescope observed this flaring activity on 2018 January 19 \citep{Kaur2018} and the reported magnitudes are $\sim 15.0$, $\sim 14.4$, $\sim 13.9$ and $\sim 13.5$ in $B$, $V$, $R$ and $I$, respectively. We compared the $B$ and $V$ magnitudes obtained from {\it Swift} UVOT observations performed on 2018 January 17 and 19 with REM and SARA-KPNO measurements, and they are consistent with each other.

\section{Temporal Analysis}\label{sec:temp_analysis}
 In 2018 January, 3C\,279 was again reported in the highest flux states at $\gamma$-ray energy  (above 100 MeV) for the first time since 2015 Junes flare, based on the detections from \emph{Fermi} \citep{Pfesesani2018} and \emph{AGILE} \citep{Lucarelli2018}. Earlier, it had been detected in active state by \emph{AGILE} during the period between 2017 December, 28-30 \citep{Pittori2017}. We carried a detailed multi-wavelength study of 2018 January flaring of 3C\,279 during the period between MJD\,58129 to 58152 using the simultaneous observations from \emph{Fermi}-LAT, \emph{Swift}-XRT, 
and \emph{Swift}-UVOT. The multi-wavelength light curve (MLC) obtained is shown in Figure \ref{fig:mwl}. The $\gamma$-ray light curve
points (top panel) are one-day binned, whereas X-ray-UV/optical flux points are per observation IDs. 
The MLC indicates substantial variation in flux in all the bands.  The $\gamma$-ray light 
curve shows a rise in flux from the quiescent flux level of $\rm (1.86\pm0.25)\times 10^{-6}\,photons\,cm^{-2}\,s^{-1}$ at MJD\,58129.5 to a peak flux 
level of $\rm (2.05\pm0.06)\times 10^{-5}\,photons\,cm^{-2}\,s^{-1}$ at MJD\,58136.5.  After the peak, the 
$\gamma$-ray flux decreases abruptly in the next two days, but before reaching the quiescent 
flux level again it stayed in plateau state for nearly three days. Later, a very high-energy $\gamma$-ray 
detection has been reported by the \emph{H.E.S.S} observations for the source during 2018 January 27-28 \citep[MJD 58145-58146;][]{2018ATel11239....1N}, which is shown by the gray 
region. However, at \emph{Fermi}-LAT energy, the $\gamma$-ray light curve did not show substantial 
flux enhancement during this period.  We also computed the highest energy of $\gamma$-ray event with high probability of being associated with the source by using the tool \emph{gtsrcprob}. The photon with highest energy of  92.56 GeV was detected on MJD 58135.09459131 (2018 January 17 02:16:12.689) with 99.98\% probability of it's origin being from 3C\,279.

\begin{figure}
\centering
\includegraphics[width=0.5\textwidth,angle=270]{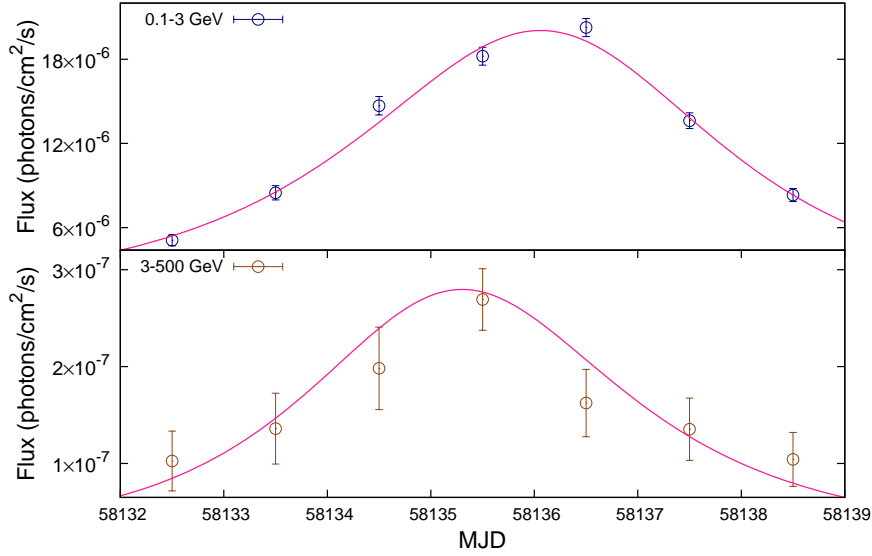}
\caption{  One-day binned $\gamma$-ray light curve obtained around the flaring period, MJD\,58132--58139, fitted with temporal profile function (Equation \ref{eq:tpf}). The top panel corresponds to energy range 0.1--3 GeV, while bottom panel represent the light curve in the 3--500 GeV energy range.}
\label{fig:tpf}
\end{figure}

\begin{figure}
\centering
\includegraphics[width=0.5\textwidth,angle=270]{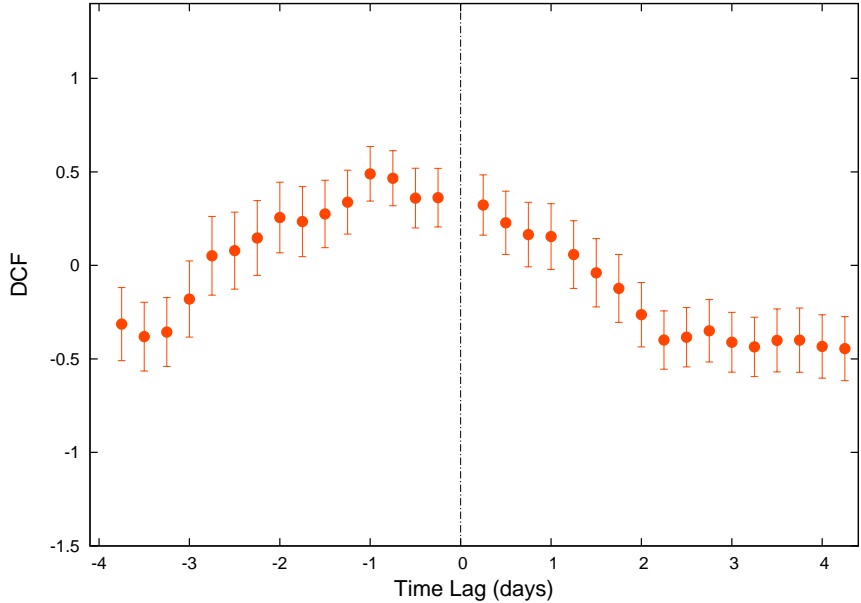}
\caption{ DCF calculated using the 6-hour binned $\gamma$-ray light curve in the low (0.1--3 GeV) and high (3--500 GeV) energy (see text for more details).}
\label{fig:dcf}
\end{figure}

Due to continuous monitoring by \emph{Fermi}-LAT, the temporal profile of the flare is quite evident in 
$\gamma$-ray band rather than X-ray or Optical/UV energies.
Therefore, to quantitatively characterize the asymmetry in the rise and falling 
time of the flare, we use the $\gamma$-ray light curve and split the one-day binned $\gamma$-ray flux
into two energy regimes: 0.1--3 GeV and 3--500 GeV. 
The 0.1--3 GeV spectra showed significant curvature and hence the flux was obtained by fitting the spectra
with a log-parabola model defined by
\begin{align}
\frac{dN(E)}{dE}=N_0\left(\frac{E}{E_0}\right)^{-\alpha-\beta\ln(E/E_0)}
\end{align}
Here, $N_0$ is the normalization, $\alpha$ is photon spectral index at pivot energy $E_0$ ($\approx$ 342 MeV) and $\beta$ is 
parameter deciding the spectral curvature. On the other hand, the curvature is negligible for the spectra beyond 
3 GeV and the flux is obtained by fitting the spectra with a simple power law function. 
The one-day binned $\gamma$-ray light curves around the flaring period, MJD\,58132--58139, in these energy 
ranges are shown in Figure \ref{fig:tpf}. 
 In order to measure the rise and fall time of 
the flare, we fitted the $\gamma$-ray light curve 
with a time profile
\begin{equation}\label{eq:tpf}
\rm F(t)=F_b + \frac{F_p}{e^{(t_p-t)/\tau_{rise}}+e^{(t-t_p)/\tau_{fall}}}
\end{equation}
where,  $F_b$ is the constant level flux, $\rm t_p$ is the time corresponding to the peak 
flux $F_p$ of flare, 
$\rm \tau_{rise}$ and $\rm \tau_{fall}$ are the characteristic rise and decay timescales 
of the light curve. 
The knowledge of $\rm \tau_{rise}$ and $\rm \tau_{fall}$ let us to estimate the total flare duration as 
$\tau_{flare}=2({\tau_{rise}+\tau_{fall}})$ \citep{Abdo2010c}.  We obtained $F_b$ by fitting a constant line to low flux points during MJD\,58121--58130. The obtained values are $\rm (1.63\pm0.18)\times10^{-6}\,ph\,cm^{-2}\,s^{-1}$ and $(\rm 2.8\pm0.7)\times10^{-8}\,ph\,cm^{-2}\,s^{-1}$ for $\rm 0.1-3\,GeV$ and $\rm 3-500\,GeV$, respectively. 
Using these values in temporal profile equation (\ref{eq:tpf}), the light curves are fitted and
the best-fit parameters are listed in Table \ref{table:tpf}. The fitted temporal profile 
is shown as solid maroon line in Figure \ref{fig:tpf}. The fit profile suggests considerable
shift in $t_p$ ($\sim 1.0$\,d) with 0.1--3 GeV light curve peaking at MJD\,58136.17; while 
for 3--500 GeV, the peak flux occurs at MJD\,58135.2. 
 In order to quantify the shift in maximum flux between between 0.1--3 GeV and 3--500 GeV $\gamma$-ray light curves,  we used the  Z-transform discrete cross-correlation function algorithm (ZDCF; \citealp{Alexander1997}) with 100 Monte Carlo draws. The one-day binned light curve has less than 11 points around the flaring, however Z-transform convergence needs at least 11 points per bin. Therefore to secure more flux points for ZDCF analysis, we obtained six-hour binned light curves in the energy range 0.1--3 GeV and 3--500 GeV during the period MJD 58131--58141 using the maximum unbinned likelihood analysis. 
By choosing 22 points per bin, the DCF curve obtained between the two energy range is shown in Figure \ref{fig:dcf}. From the ZDCF, we obtained a time lag of $-1.0_{+0.63}^{-0.31}$ day (the uncertainties are at $1\sigma$ confidence level) between the low (0.1--3.0 GeV) and high (3.0--500 GeV) energy $\gamma$-ray emission (an analogous result is found by choosing 0.1--1 GeV and 1--300 GeV energy band light curves). 
It is interesting to note that, a similar time lag of one day between the maximum optical polarization and the peak optical/$\gamma$-ray flux has been observed in the follow-up optical observations of the 2015 June $\gamma$-ray flare with GASP-WEBT \citep{Pittori2018}. This resemblance suggest that  
the lag at low and high energy $\gamma$-ray emission is possibly related to the behavior of intrinsic alignment of magnetic field in the jets. 
Further, the light curves are asymmetric with a slow rise--fast decay trend observed for 0.1--3 GeV energy range 
and fast rise--slow decay trend observed for 3--500 GeV energy range. If we attribute this asymmetry
to the difference in the timescales associated with the strengthening and weakening of the underlying acceleration 
process, then one would observe a similar trend in both the light curves. However, the dissimilar trends observed in
these light curves indicate an additional process manifesting the flare. A plausible reason may be associated with the shift 
in SED peak energy during the flare. A log-parabola representation of the 0.1--3 GeV spectra indicates that this 
energy regime may fall
close to the SED peak; whereas a power law representation of 3--500 GeV spectra indicates the spectral regime
well beyond the SED peak. If this inference is correct then the high energy light curve indicates the development and decay of the 
acceleration process while the low energy light curve may be additionally influenced by the shift in SED peak.

\begin{table*}
	\centering
\setlength{\tabcolsep}{18.0pt}
	\begin{tabular}{lccccr}
		\hline
		Energy bin (GeV) & Peak time of flux (in MJD) & $\rm \tau_{rise}$ & $\rm \tau_{fall}$ & $\tau_{flare}$ & $\chi^2$/d.o.f \\
		\hline

		0.1--3 & 58136.17 & 1.64$\pm$0.06 & 1.42$\pm$0.06 & 6.12$\pm$0.08 & 7.33/5 \\
	        3--500 & 58135.2 & 1.24$\pm$0.18 & 1.45$\pm$0.17 & 5.36$\pm$0.25 & 3.90/5\\
		\hline
	\end{tabular}
	\caption{Best-fit parameter obtained by fitting the individual $\gamma$-ray light curves. Col:- 1: energy bin, 2: time corresponding to the approximate peak of the flare, 3: rise time of light curve, 3: decay time of light curve, 4: total duration of the flare, 5: $\chi^2$ and degrees of freedom.}
\label{table:tpf}
\end{table*}

\begin{figure}
\centering
\subfigure{\includegraphics[scale=0.65,angle=0]{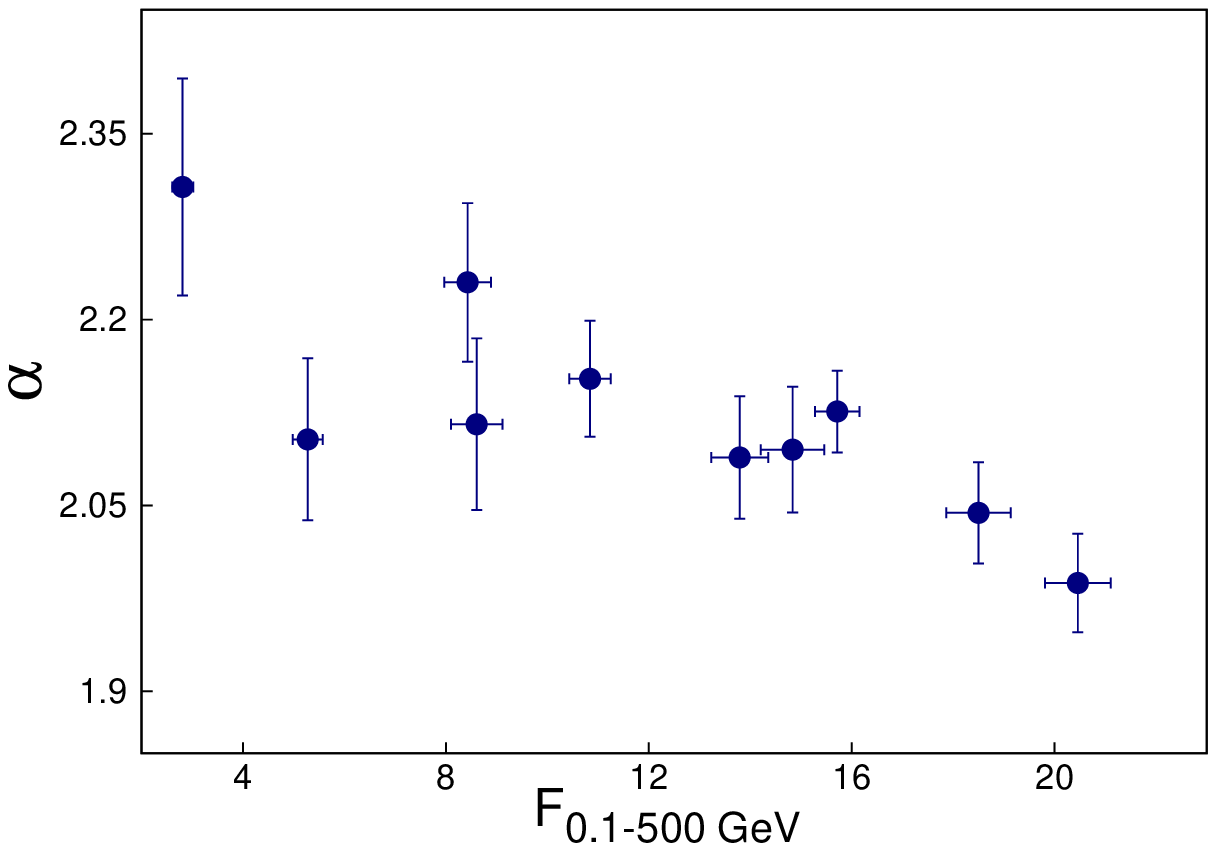}}\quad \hspace{-0.5cm}
\subfigure{\includegraphics[scale=0.65,angle=0]{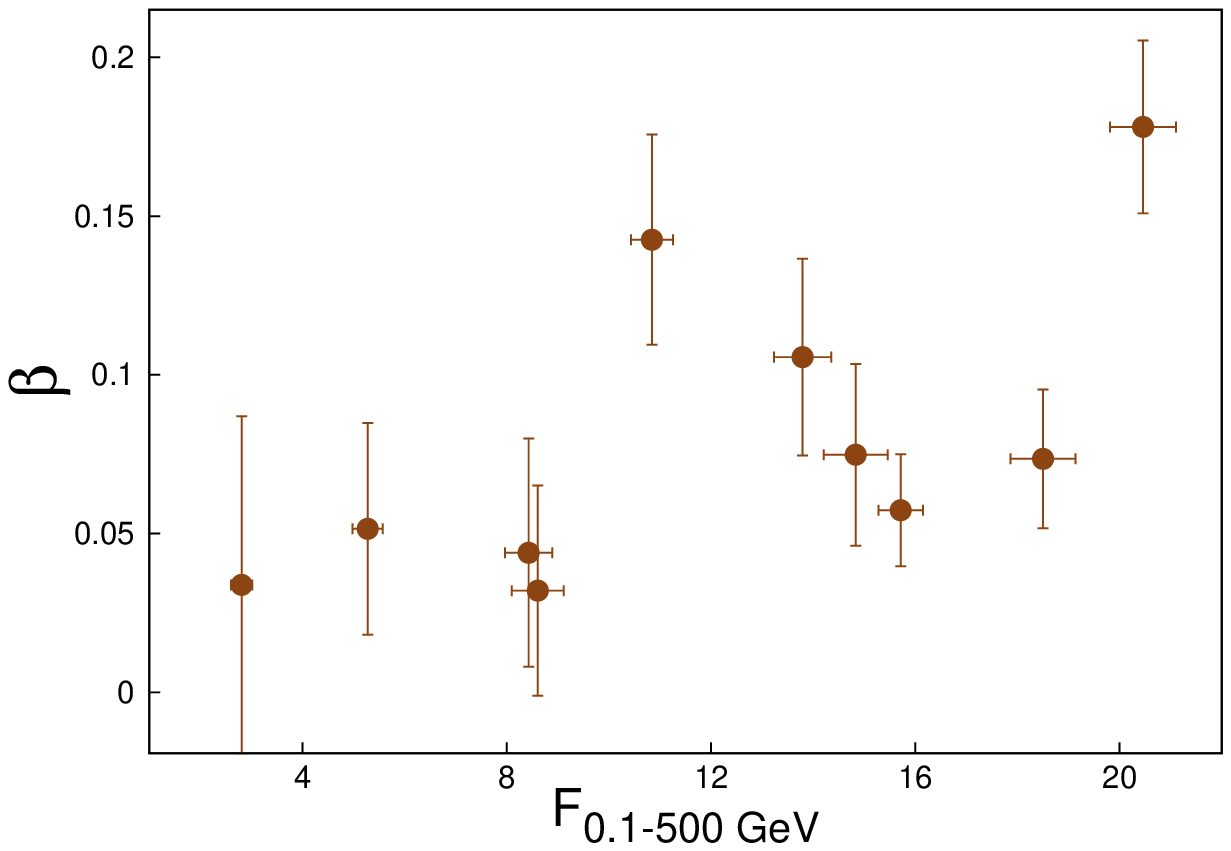}}\\
\subfigure{\includegraphics[scale=0.65,angle=0]{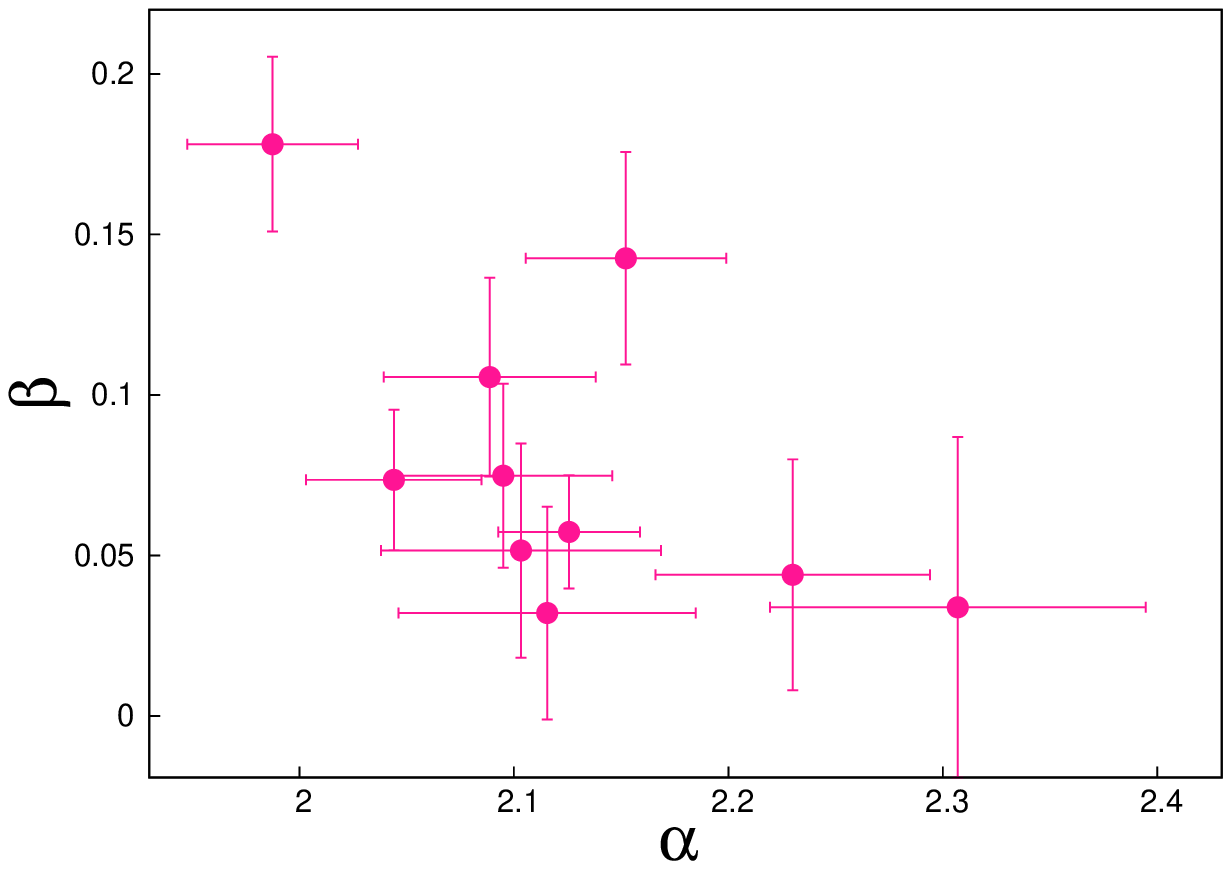}}\quad \hspace{-0.5cm}
\subfigure{\includegraphics[scale=0.65,angle=0]{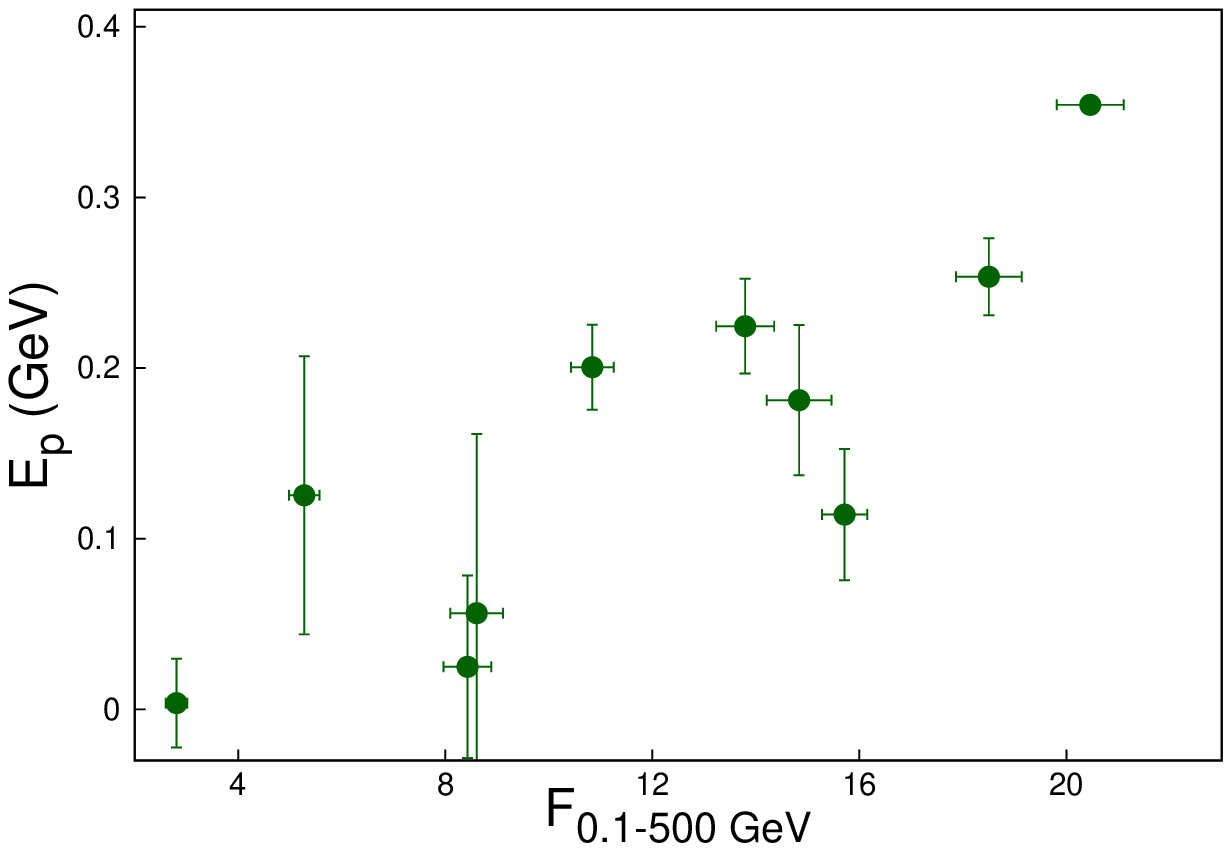}}\\
\caption{The scatter plots between one-day binned $\gamma$-ray flux distribution in the energy range 0.1--500 GeV and log-parabola model fit spectral parameters ($\alpha$ and $\beta$). Top panel left; flux (in units of $\rm 10^{-6}\, photons\,cm^{-2}\,s^{-1}$) vs $\alpha$; \quad top panel right: flux (in units of $\rm 10^{-6} photons\,cm^{-2}\,s^{-1}$) vs $\beta$; \quad bottom panel left : $\alpha$ vs $\beta$; and bottom panel right : flux (in units of $\rm 10^{-6} photons\,cm^{-2}\,s^{-1}$) vs peak energy $E_p$. }
\label{fig:corelation}
\end{figure}

\begin{figure}
\centering
\includegraphics[width=0.7\textwidth,angle=0]{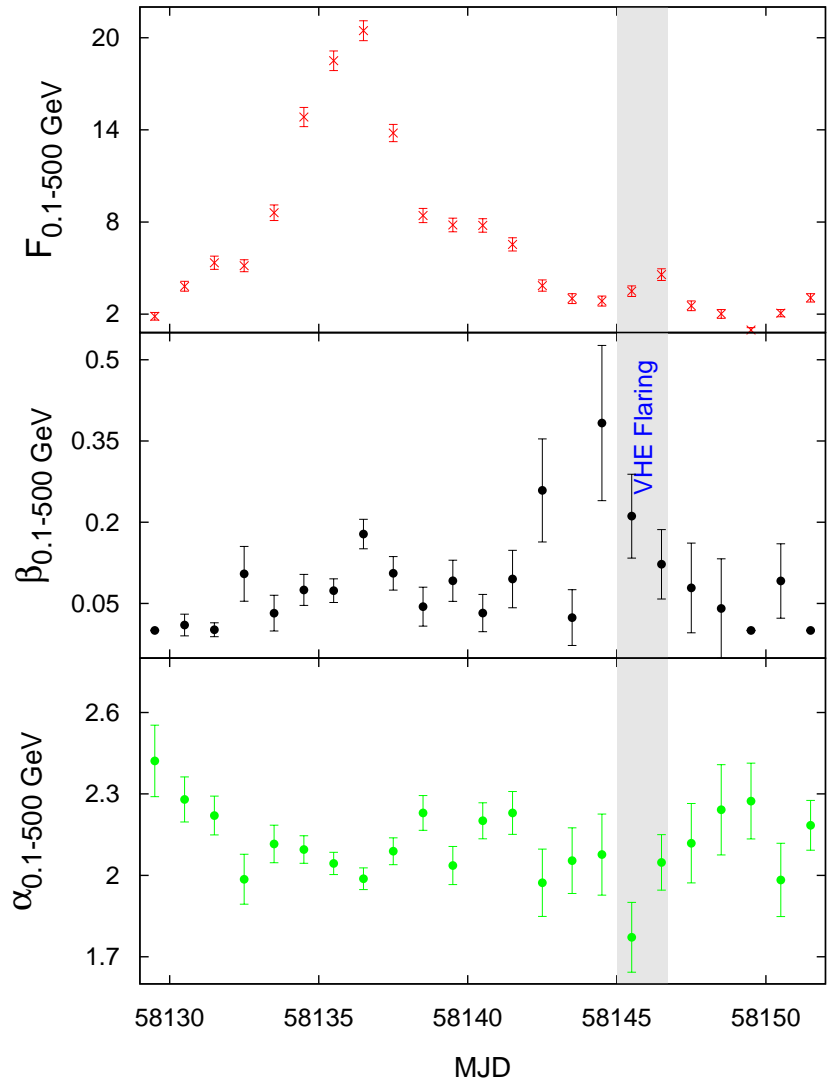}
\caption{Top panel: one-day binned $\gamma$-ray light curve (flux in units of $10^{-6}\rm\,ph\,cm^{-2}\,s^{-1}$) in the energy range 0.1--500 GeV obtained during the time period MJD\,58129--58152; middle panel: the variation in curvature parameter of the log-parabola fit to spectra with time and bottom panel: corresponding variations of spectral parameter of log-parabola fit with time.}
\label{fig:index_var}
\end{figure}

To further diagnose the effect of peak shift in 0.1--3 GeV spectra during the flare,
 we perform a Spearman rank correlation test \citep{Spearman1904} between the best-fit log-parabola parameters
over the flaring period 58130.5 -- 58141.5 MJD. We obtained the peak energy of the Compton component of the SED
from  log-parabola parameters using a relation
\begin{equation}
E_p=E_0\exp\left(\frac{2-\alpha}{2\beta}\right)
\end{equation}
  The scatter plot between $\rm F_{0.1-500\, GeV}$, $\alpha$, $\beta$ and $\rm E_p$ are shown in Figure \ref{fig:corelation} 
and the correlation study results are shown in Table \ref{table:spearman}. It is noted that the uncertainty on the curvature values are not well constrained below the flux level of $\rm 8\times10^{-6} ph\,cm^{-2}\,s^{-1}$ in the one-day binning of $\gamma$-ray data. Hence, in such cases we binned the data over two days. Significant negative correlation is observed between 
$\alpha$ and $\beta$ with correlation coefficient $\rho =-0.58$ and null hypothesis probability $\rm P_{rs} = 7.38\times10^{-2}$.
 Since around the SED peak the curvature is expected to be maximum with a hard spectra and the \emph{Fermi} energy range lie on and beyond the SED 
peak of 3C\,279, this increase in curvature  with spectral hardening  support the hypothesis of a  shift in the SED peak during the flare. 
The negative correlation observed between $\alpha$ and $\rm F_{0.1-500\, GeV}$, with $\rho = -0.75$ and $\rm P_{rs} = 1.33\times10^{-2}$, suggests 
the spectral hardening during high flux and this again indicates the high energy shift of the SED peak during high flux state. 
A nominal positive correlation is observed between $\beta$ and $\rm F_{0.1-500\,Gev}$, with $\rho = 0.68$ and $\rm P_{rs} = 2.88\times10^{-2}$,
which also confirms the high energy peak shift during high flux. 
 This inference was further confirmed directly by the positive correlation obtained between the 
$\rm F_{0.1-500\, GeV}$ and $E_p$ with $\rho = 0.78$ and $\rm P_{rs} = 7.54\times10^{-3}$, as shown in Figure \ref{fig:corelation}, bottom panel right. Though at low 
flux states $E_p$ falls beyond the energy range considered here, in comparison with other correlation results
it is evident that the rise and decay time of $F_{0.1-500\,\rm GeV}$ light curve is manifested by the shift in the SED peak.
Interestingly, from the temporal evolution of $\alpha$ and $\beta$ over the entire duration considered here (see Figure \ref{fig:index_var}), it is evident that the curvature was large ($0.21\pm0.07$) with hardest spectra ($1.77\pm 0.13$) during the epoch associated with the VHE flare. This possibly indicates a large high energy shift in SED peak 
during this period and there by enabling a significant enhancement in the VHE emission. 

To compare the $\gamma$-ray variability of 3C\,279 with the other wavebands, we calculated the 
fractional variability as \citep{Rodriguez1997, Vaughan2003}.
\begin{equation}\label{eq:fvar}
\rm F_{var}=\sqrt{\frac{S^2-\Delta^2}{<f>^2}}
\end{equation}
Here $S^2$ is variance, $<f>$ is unweighted mean flux and $\Delta^2$ is mean square value of uncertainties. 
The estimated $\rm F_{var}$ 
for $\gamma$-ray, X-ray and optical/UV energies are given in Table \ref{table:var}.  The variability amplitude at optical/UV energies is smaller ($\rm F_{var} \sim $ 0.22--0.25), while there is a substantial increase in the variability amplitude at X-ray energies (0.3--10 keV) with $\rm F_{var}=$ 0.53 and strong variation at $\gamma$-ray energies with $\rm F_{var}=$ 0.80. Thus, the variability at high energy is large compared to the low energies, which is consistent with other blazars \citep[see e.g., ][]{Zhang2005, Vercellone2010}. Since the cooling time of the high energy electrons is much shorter than that of the low-energy electrons, the large amplitude variations at $\gamma$-rays suggest that these variations comes from the high energy electrons, while small variations comes from the low energy tail of the electron distribution. The increase in variability amplitude with energy is also interpreted as the signature of spectral variability \citep{Zhang2005}. The energy dependence of variability can be associated with the hardening of the source spectrum as it becomes brighter \citep{Rodriguez1997}. In addition to large variability amplitude at $\gamma$-rays, the flaring  is also associated with Compton dominance i.e,  there is $\sim 8.5$ time increase in $\gamma$-ray flux from QS to FS, while the maximum increase in optical/UV flux from QS to FS is $\sim 2.5$ (see Table \ref{table:lat} and \ref{table:uvot}). In the standard one zone model, the $\gamma$-ray emission in FSRQs is dominated by the inverse Compton scattering of external target photon field \citep{Sahayanathan2012,Shah2017}. Therefore, the Compton dominance can be associated with the increase in bulk Lorentz factor ($\Gamma$) of the emission region, which enhances the target photon energy density by $\Gamma^2$ in the rest frame of emission region.
\begin{table*}
	\centering
\setlength{\tabcolsep}{18.0pt}
	\begin{tabular}{@{}lcc}
		\hline
        & \multicolumn{2}{c}{Spearman correlation parameters} \\ \cline{2-3}
		Correlation & $\rho$ & $\rm P_{rs}$ \\
		\hline
		\vspace{1.5mm}
		Flux vs $\alpha$ & -0.75 & $1.33\times10^{-2}$ \\
		\vspace{1.5mm}
		Flux vs $\beta$ &  0.68 & $2.88\times10^{-2}$	\\
		\vspace{1.5mm}
		Flux vs $\rm E_p$ &  0.78 & $7.54\times10^{-3}$ \\
		\vspace{1.5mm}
		$\alpha$ vs $\beta$ & -0.58 & $7.38\times10^{-2}$ \\		
		\hline
	\end{tabular}
	\caption{Spearman Correlation results obtained by comparing one-day binned $\gamma$-ray flux distribution and log-parabola model fit parameters ($\alpha$ and $\beta$) in the energy range 0.1-500 GeV during the period 58130.5-58141.5.}
\label{table:spearman}
\end{table*} 

\begin{table}
	\centering
\setlength{\tabcolsep}{35.0pt}
	\begin{tabular}{lr}
		\hline
		Energy Band & $\rm F_{var}$ \\
		\hline
		
		$\gamma$-ray(0.1--500 GeV) & 0.801 \\
		X-ray (0.3-10 KeV) & 0.532\\
		M2 & 0.249\\
		W2 & 0.247\\
		W1 & 0.221\\
		B  & 0.234\\
		U  & 0.221\\
		\hline
	\end{tabular}
	\caption{Variability amplitude obtained for different bands of the MLC. Col:- 1: energy band, 2: variability amplitude.}
	\label{table:var}
\end{table}

\section{Spectral Analysis}\label{sec:spec_analysis}

To investigate the spectral properties and the source behavior during different flux states, we chose three time domains
where simultaneous observations in $\gamma$-ray, X-ray and optical/UV energies were available. The three flux states are categorized as 
flaring state (FS: MJD\,58134--58138), plateau state (PS: MJD\,58138--58142) and quiescent state (QS: MJD\,58150.2--58151.2) which are indicated by 
vertical dotted lines in Figure \ref{fig:mwl}. The $\gamma$-ray data during these states are well reproduced by a log-parabola function and the 
resultant time averaged flux along with the best-fit parameters are summarized in Table \ref{table:lat}. 
To obtain the X-ray flux, we added the individual spectra in FS, PS and QS using the {\sc ftool addascaspec} and fitted with an absorbed 
power law model. The best-fit parameters and unabsorbed flux for the three states are listed in Table \ref{table:xrt}. In the case 
of {\it Swift}-UVOT, we added the individual images in three different flux states using the {\sc uvotimsum} task and derive the 
UV/optical flux values from the combined image, which are provided in Table \ref{table:uvot}.

\begin{table*}
\setlength{\tabcolsep}{12.0pt}
	\begin{tabular}{@{}lcccccc}
		\hline
		Flux state & Time period (MJD) & Flux (0.1--500 GeV) & $\alpha$ & $\beta$ & Norm. & TS\\
		\hline
		FS & 58134-58138 & 17.09$\pm$0.31 & 2.05$\pm$0.02 & 0.11$\pm$0.03 & 157.82 & 20757\\
		PS & 58138-58142 & 7.71$\pm$0.24 & 2.17$\pm$0.04 & 0.06$\pm$0.02 & 65.44 & 7758 \\
		QS & 58150.2--58151.2  & 2.03$\pm$0.24 & 1.96$\pm$0.14 & 0.10$\pm$0.06 & 19.10 & 496 \\	
		\hline
	\end{tabular}
	\caption{Summary of log-parabola spectral fit to \emph{Fermi}-LAT observations in three flux states. Col:- 1: flux state, 2: time period (MJD), 3: 0.1--500 GeV integrated $\gamma$-ray flux in units of $10^{-6}\rm\,ph\,cm^{-2}\,s^{-1}$, 4,5: spectral parameters of log-parabola model, 6: normalization in units of $10^{-10}$ and 7: test statistics.}
	\label{table:lat}
\end{table*} 

\begin{table*}
\setlength{\tabcolsep}{12.0pt}
\begin{tabular}{@{}lcccccc}
\hline
Flux state & V & B & U & UVW1 & UVM2 & UVW2\\
\hline
FS  & $6.90\pm 0.17$ & $6.53\pm 0.15$ & $8.38\pm 0.21$ & $10.09\pm 0.32$ & $12.68\pm 0.26$ & $12.01\pm 0.30$ \\
PS  & $6.27\pm 0.17$ & $7.22\pm 0.15$ & $7.98\pm 0.21$ & $9.29\pm 0.31$ & $11.11\pm 0.25$ & $11.02\pm 0.28$\\
QS  & $2.94\pm 0.11$ & $3.38\pm 0.09$ & $3.71\pm 0.11$ & $4.43\pm 0.16$ & $4.96\pm 0.13$ & $5.04\pm 0.14$\\
\hline
\end{tabular}
\caption{Observed flux values of 3C\,279 obtained from \emph{Swift}-UVOT. Col.:- 1: flux states, 2-7: extinction corrected Flux at V, B, U, UVW1, UVM2 and UVW2 bands in units of $\rm 10^{-15} erg\,cm^{-2}\,s^{-1}\,\AA^{-1}$.}
\label{table:uvot}
\end{table*}

\begin{table}
		\centering
		\begin{tabular}{@{}lccccr@{}}
			\hline
			  Flux & Exposure & PL Index & Flux ($0.3-10$ keV)  & $\chi^2$/d.o.f \\
                   state & time (s) & ($\Gamma_{X}$) & ($\rm 10^{-11}\ erg~cm^{-2}~s^{-1}$) &  \\
			\hline
              FS & 2415 & $1.28^{+0.05}_{-0.05}$ & $8.06^{+0.44}_{-0.44}$ & $91.2/76$ \\
              PS & 2051 & $1.57^{+0.06}_{-0.06}$ & $4.57^{+0.24}_{-0.23}$ & $89.8/92$ \\
              QS & 1735 & $1.54^{+0.09}_{-0.09}$ & $1.96^{+0.17}_{-0.16}$ & $31.4/32$ \\
			\hline
		\end{tabular}
		\caption{Best-fit X-ray spectral parameters of 3C\,279 in the three flux states when fitted with PL. Col:- 1: flux state, 2: exposure time in seconds, 3: power law index, 4: flux in $0.3-10$ keV band, 5: $\chi^2$ and degrees of freedom.}
		\label{table:xrt}
\end{table}

For each state, the $\gamma$-ray SED points are obtained by dividing the total energy (0.1--500 GeV) into 10 energy bins. Assuming the contribution of sources in the ROI, other than 3C\,279, does not change with energy, we freeze the parameters of these sources to their best-fit values obtained in the energy range 0.1--500 GeV and performed the unbinned likelihood analysis in each bin.

\begin{figure}
	\centering
	\subfigure{\includegraphics[scale=0.32,angle=270]{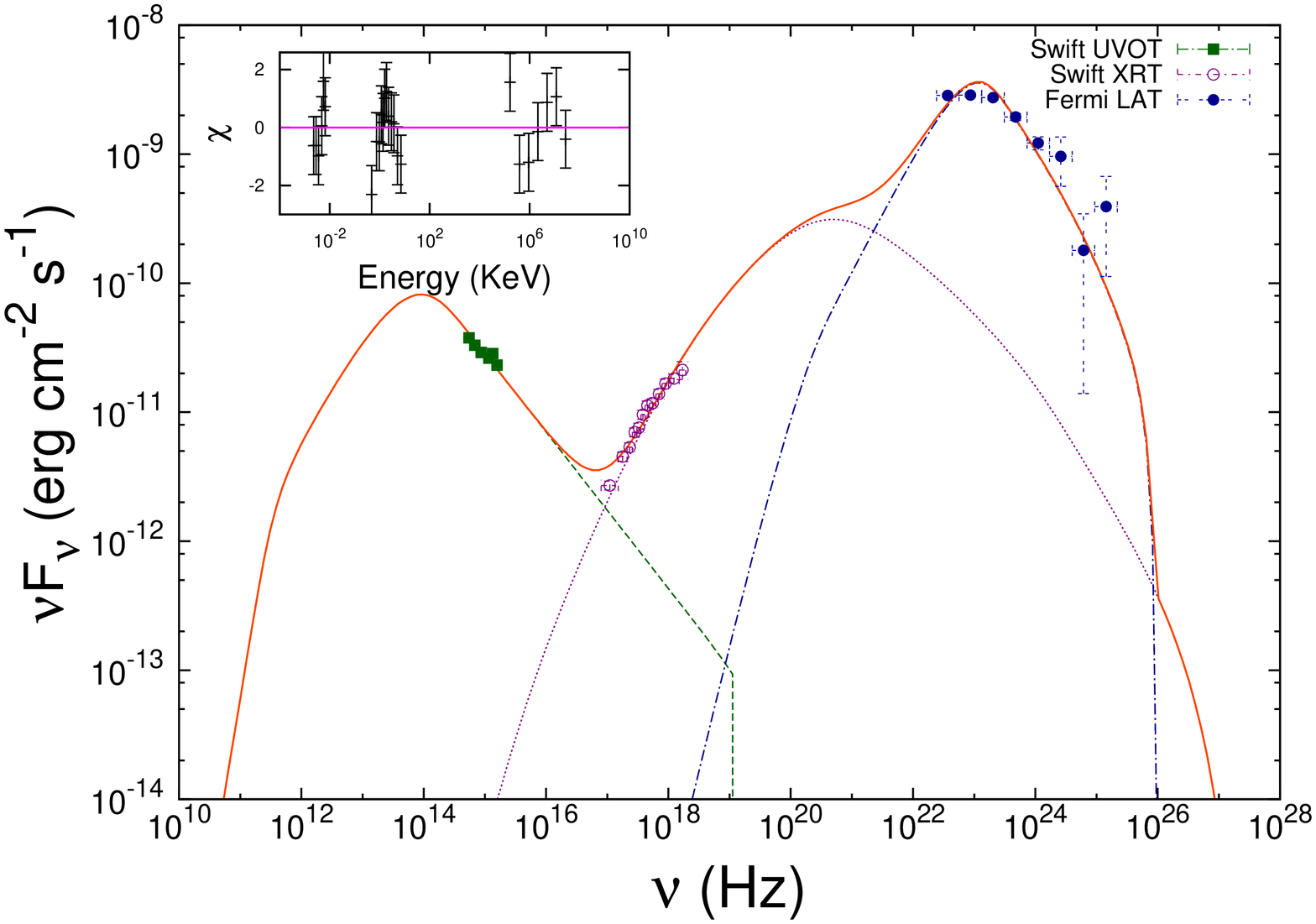}}\quad\hspace{-0.4cm}
	\subfigure{\includegraphics[scale=0.32,angle=270]{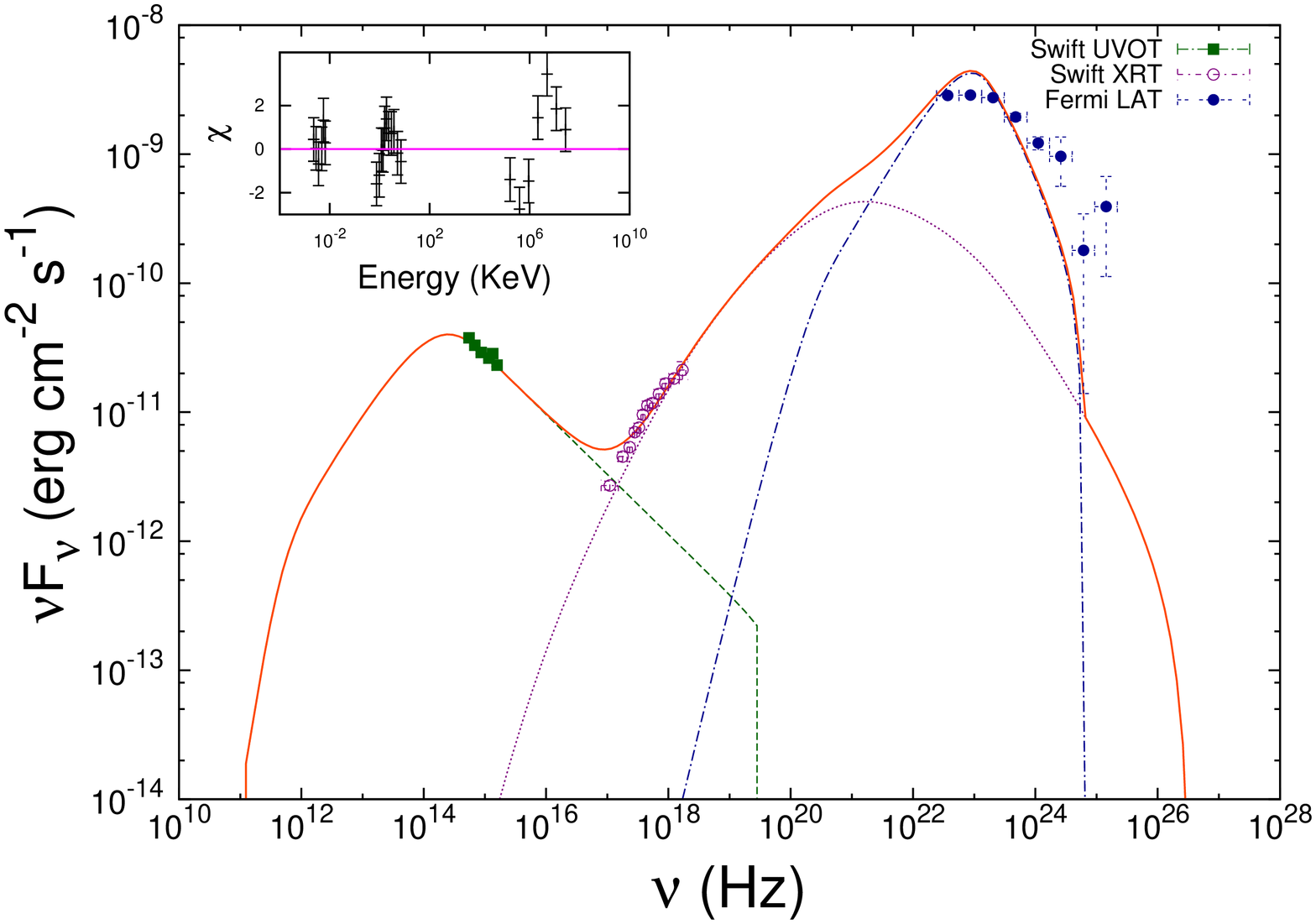}}\\
     \caption{SED of 3C\,279 obtained during the flaring state (MJD\,58134--58138) by considering target photons from IR region of temperature 1000 K ($left$) and BLR region ($right$). The observed flux points are specified by filled squares (\emph{Swift}-UVOT), open circle (\emph{Swift}-XRT) and filled circles (\emph{Fermi}-LAT). The best-fit synchrotron model spectrum is represented by dashed line, SSC model spectrum by a dotted line, EC spectrum by dot-dashed line and total spectrum by a solid line. The fitting is done in XSPEC using synchrotron, SSC and EC models. The inset plots shows residual with the systematics of 12\%. In case of IR photons, the SED model provides a good fit to the observed spectrum with reduced $\chi^2$ of 1.19 for 20 d.o.f, while as in case of BLR photons the fit is poor with reduced $\chi^2$ of 2.78 for 20 d.o.f.} 
\label{fig:sed_flare.fit}
\end{figure}

\begin{figure}
	\centering
	\subfigure{\includegraphics[scale=0.32,angle=270]{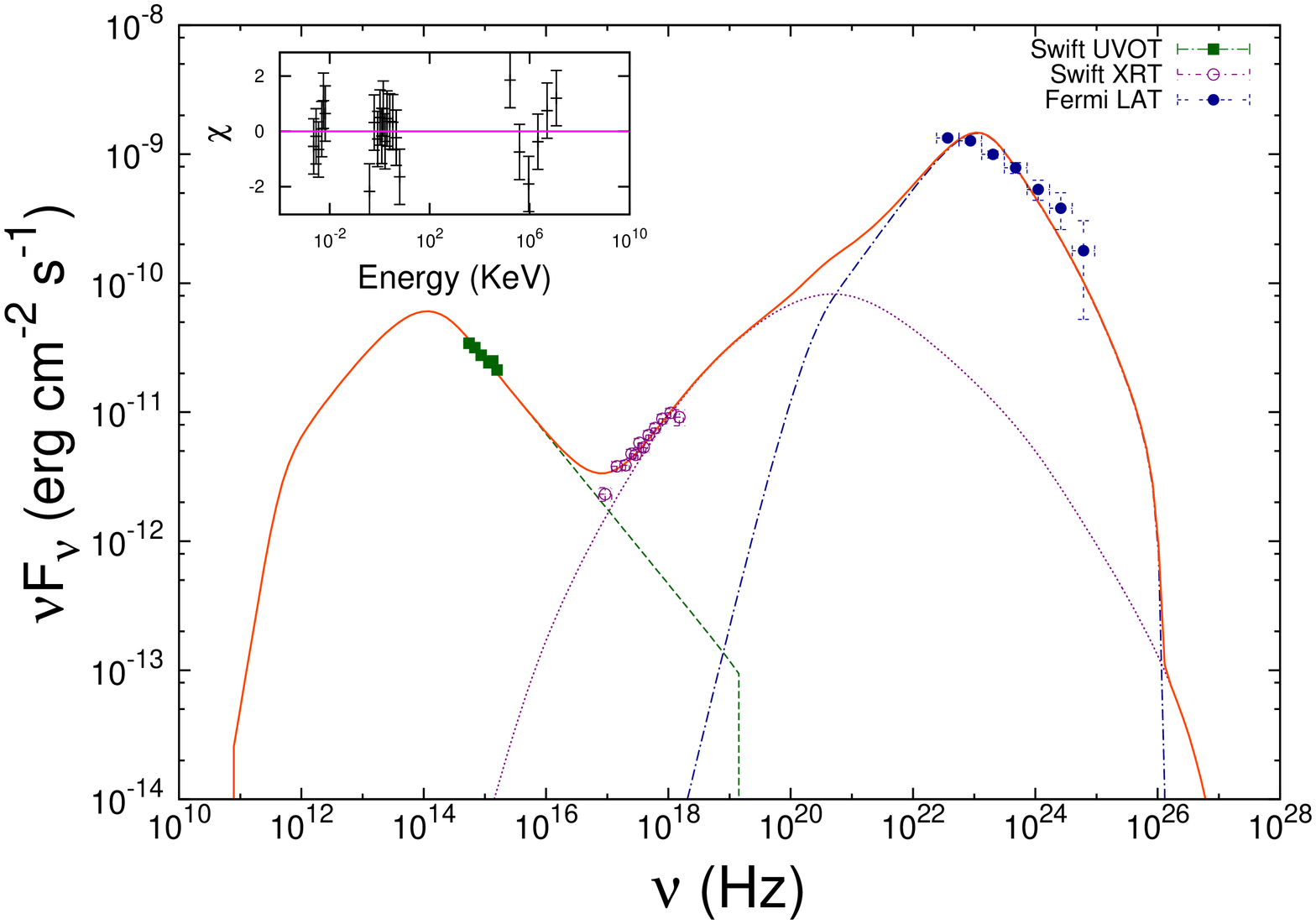}}\quad \hspace{-0.4cm}
	\subfigure{\includegraphics[scale=0.32,angle=270]{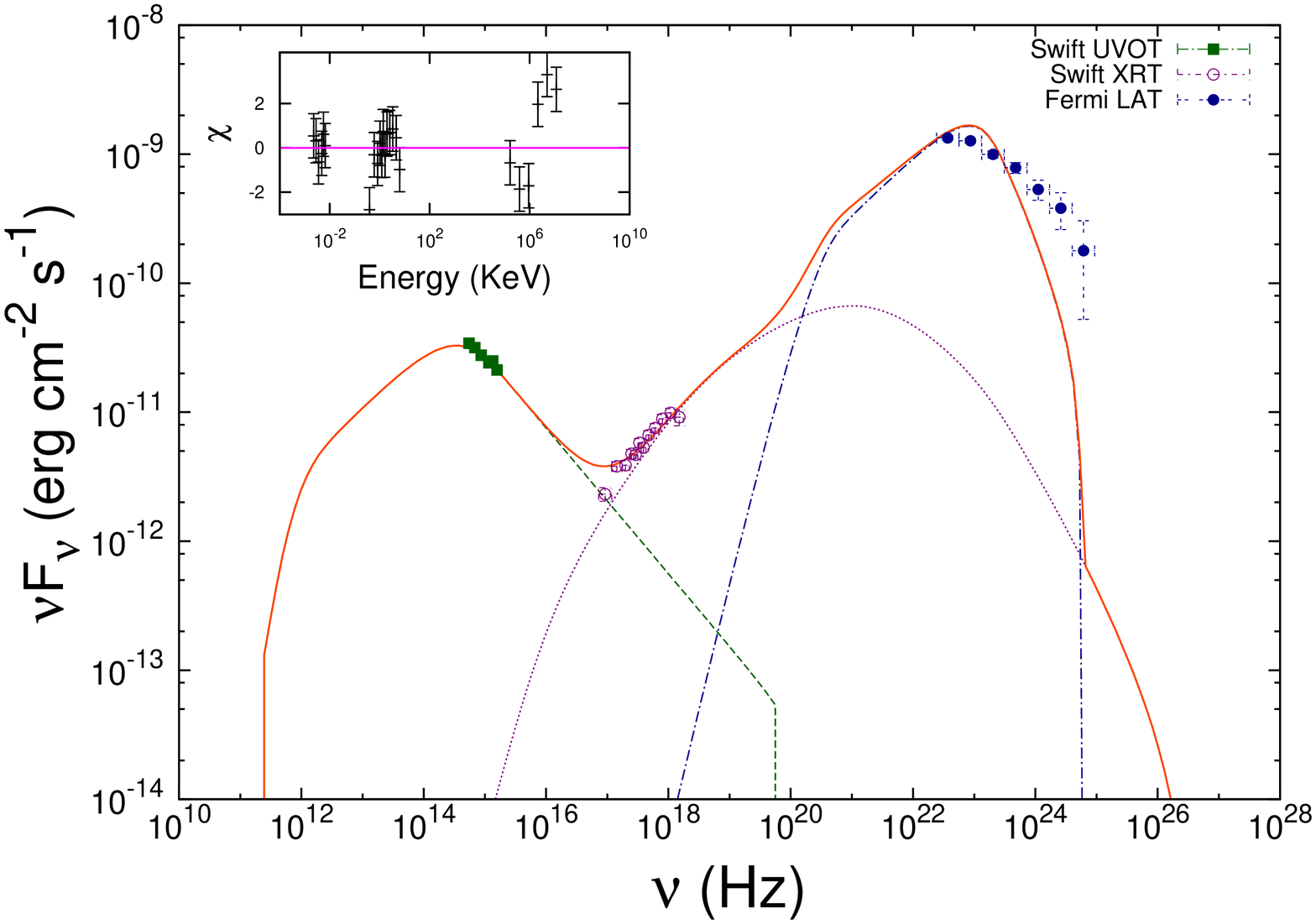}}
	\caption{SED of 3C\,279 obtained during the plateau state (MJD\,58138--58142) by considering target photons from IR region of temperature 1000 K ($left$) and BLR region ($right$). The symbols and fitting curves have the same information as in Figure \ref{fig:sed_flare.fit}. The SED model with IR target photons provides a good fit to the observed spectrum with reduced $\chi^2$ of 1.12 for 19 d.o.f, while as in case of BLR photons the fit is poor with reduced $\chi^2$ of 2.20 for 19 d.o.f.} 
\label{fig:sed_platau.fit}
\end{figure}

\begin{figure}	
     \centering
     \subfigure{\includegraphics[scale=0.32,angle=270]{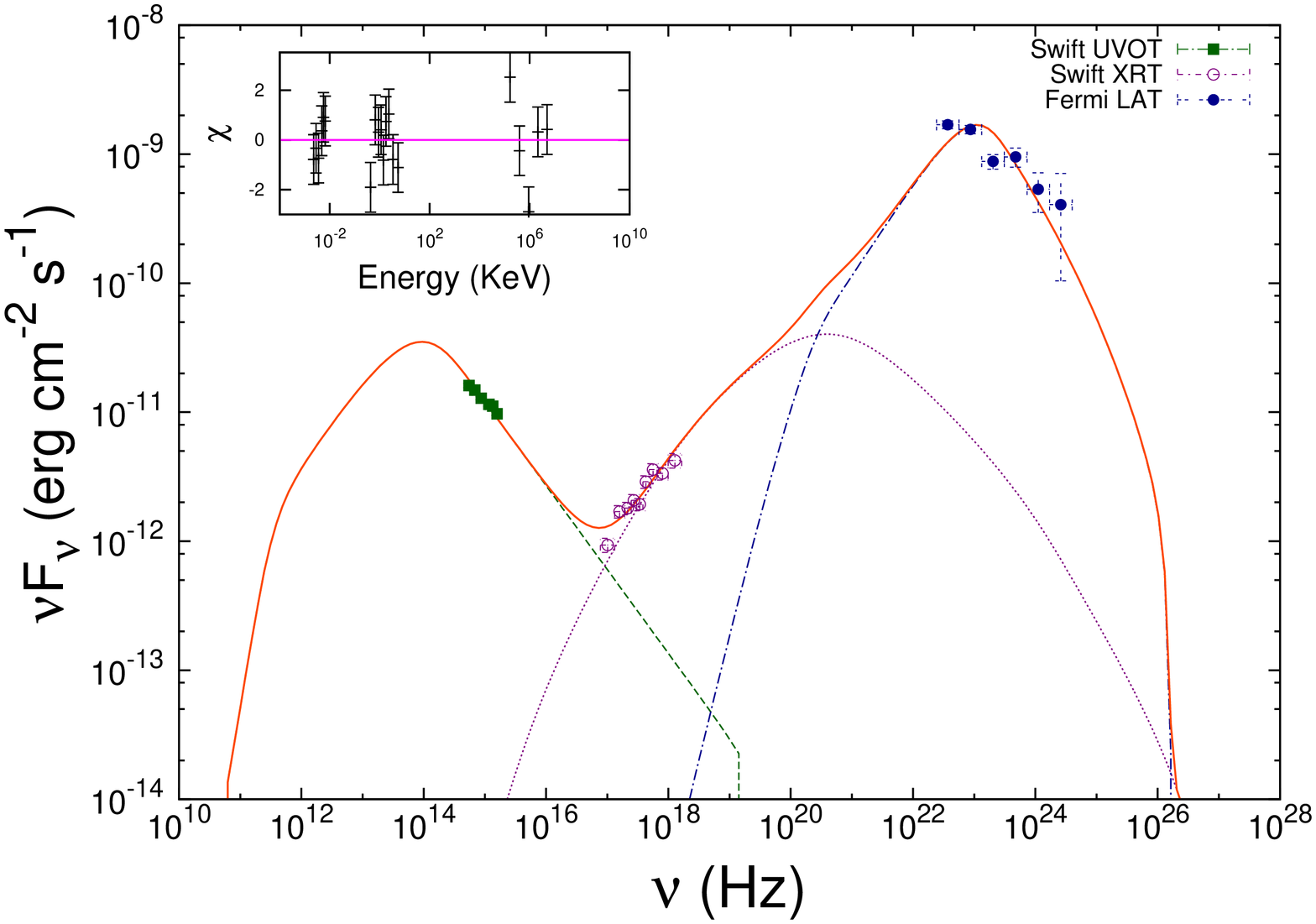}}\quad \hspace{-0.4cm}
	 \subfigure{\includegraphics[scale=0.32,angle=270]{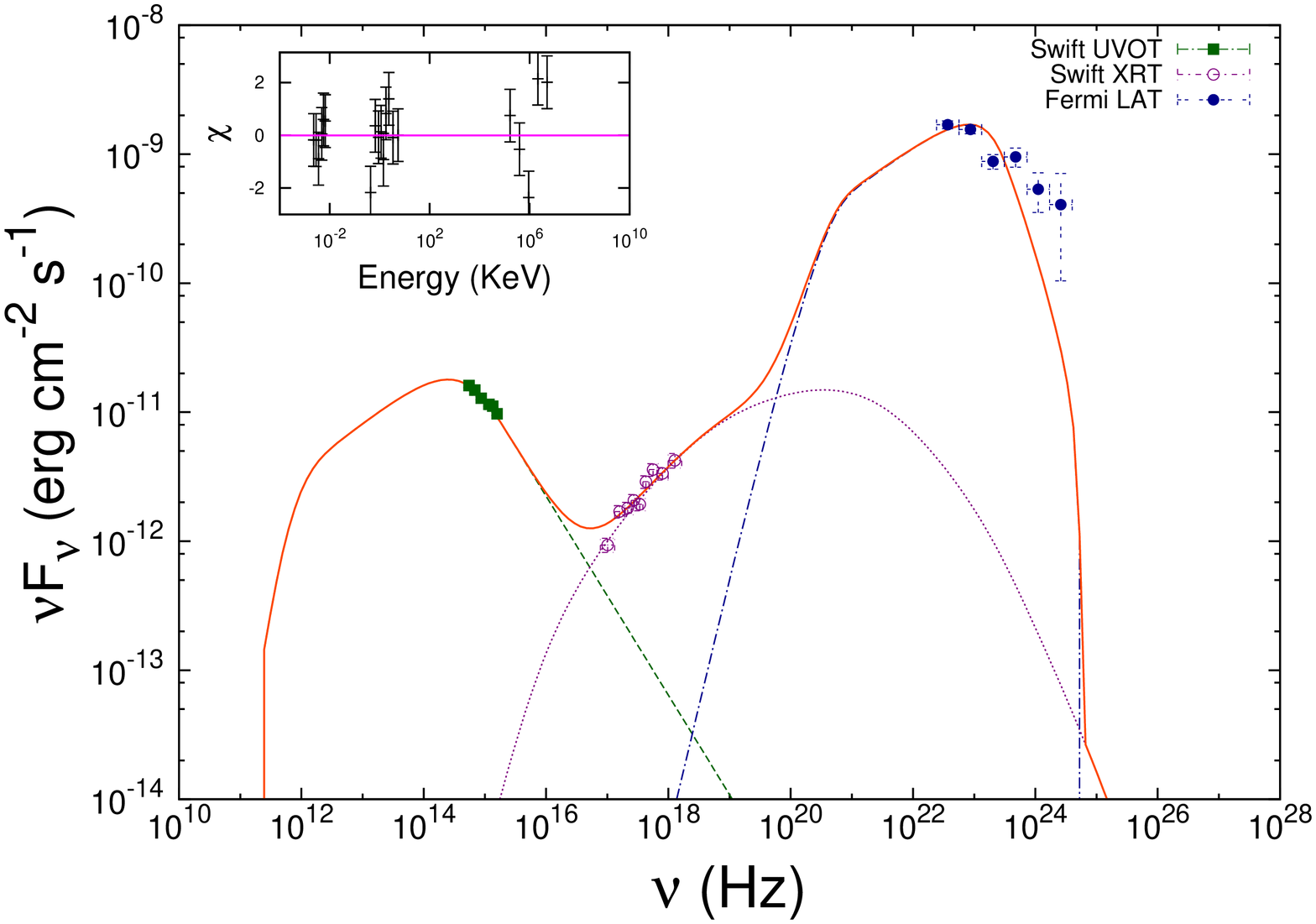}}
      \caption{SED of 3C\,279 obtained during the quiescent flux state (MJD\,58150.2--58151.2) by considering target photons from IR region of temperature 1000 K ($left$) and BLR region ($right$). Again the symbols and fitting curves have the same information as in Figure \ref{fig:sed_flare.fit}. The SED model with IR target photons provides a good fit to the observed spectrum with reduced $\chi^2$ of 1.78 for 15 d.o.f, while as in case of BLR photons the fit has reduced $\chi^2$ of 1.66 for 15 d.o.f.} 
\label{fig:sed_qs.fit}
\end{figure}

The $\gamma$-ray emission, in case of FSRQs, are mainly dominated by the IC scattering of external target photons \citep{Shah2017, Sahayanathan2018}. 
The dominant sources of external target photons are the emission from the broad line region (EC/BLR) at $\approx 2.47\times10^{15}$ Hz or the thermal IR photons 
at $\approx 1000$ K from the dusty torus (EC/IR). 
To model the broadband SED, corresponding to the flux states considered here, we assume a simple scenario where
the emission region is assumed to be a spherical plasma cloud of radius $R$ filled with a broken power-law electron distribution given by
\begin{align}
\label{eq:broken}
N(\gamma) d\gamma =\left\{
	\begin{array}{ll}
K \gamma^{-p}d\gamma,&\mbox {~$\gamma_{{\rm min}}<\gamma<\gamma_b$~} \\
K \gamma^{q-p}_b \gamma^{-q}d\gamma,&\mbox {~$\gamma_b<\gamma<\gamma_{{\rm max}}$~}
\end{array}
\right.
\end{align}
Here, $\gamma$ is the Lorentz factor of the electrons with $\gamma_{\rm min}$ and $\gamma_{\rm max}$ as the limiting values, $p$ and $q$
are the low energy and high energy electron indices, $\gamma_b$ the Lorentz factor of the electrons corresponding to the power-law
break and $K$ is the normalization. The emission region is permeated by a tangled magnetic field $B$ and move down the jet with a bulk
Lorentz factor $\Gamma$, aligned at an angle $\theta$ with respect to the line of sight of the observer. The resultant spectra corresponding to
synchrotron, SSC and EC scattering of the external photon field are estimated numerically. This numerical model is coupled as a local model in 
{\sc xspec} to perform a statistical fitting of the broadband SED corresponding to different flux states \citep{Sahayanathan2018}. 
To account for the effect of the temporal evolution
of the observed SED during the integration time and the model related uncertainties, we add a 12\% systematic error 
evenly over the entire data to obtain a reasonable reduced $\chi^2$ value. The spectral fit is performed for two cases of the
external photon field corresponding to the emission from dusty torus and BLR region. The limited information available at
optical, X-ray and $\gamma$-ray energy bands force us to fix most of the parameters to their typical values and fitting
is performed over $p$, $q$, $\Gamma$, $B$ and the electron energy density $U_e$. The best-fit parameters for the two cases 
of external photon field is given in Table \ref{table:sed} and the model spectra along with the observed SED is shown 
in Figures \ref{fig:sed_flare.fit}, \ref{fig:sed_platau.fit} and \ref{fig:sed_qs.fit} for FS, PS and QS, respectively. The SED modelling with IR target photons provides a good fit to the observed spectrum with reduced $\chi^2_{r}$ per
degrees of freedom as 1.19/20 for FS, 1.12/19 for PS and for QS we obtained 1.78/15. In case of BLR target photon field, 
the $\chi^2_r$/d.o.f is quite high with 2.78/20 for FS and 2.20/19 for PS, while its value of 1.66/15 for QS is is acceptable. Due to high $\chi^2_r$ values
the upper and lower bounds of the fit parameters are not obtained for FS and PS.

The broadband SED fitting during different flux states; FS, PS and QS, are reasonable under these emission mechanisms (Figures \ref{fig:sed_flare.fit}, \ref{fig:sed_platau.fit} and \ref{fig:sed_qs.fit}) though 
the fit parameters obtained when the $\gamma$-ray spectra is attributed to EC/IR are more acceptable than EC/BLR. Particularly, the
$\Gamma$ obtained through EC/BLR is too low and such low value can question the $\gamma$-ray opacity of the emission region 
against pair production losses \citep{Dondi1995}. It can also be noted that $U_e$ and the magnetic field 
energy density ($U_B$) are close to equipartition under both these $\gamma$-ray emission models. However, in the case of 
EC/BLR, $U_B$ exceeds $U_e$ and such high magnetic pressure may cause rapid quenching of the non-thermal electron distribution,
unless it is replenished instantly. Besides these, the Klein-Nishina decline of EC/BLR spectrum begins at relatively much lower energy 
($\approx 50$ GeV) compared to EC/IR spectrum and hence VHE detection of the source is not possible in case of the latter. 
Though the information at VHE is not available during this epoch, the later detection of 3C\,279 at VHE supports EC/IR
mechanism and the location of the emission region to be beyond the BLR region.
Both the models fail to reproduce the smooth Compton peak suggested by the \emph{Fermi} $\gamma$-
ray data. This can be possibly associated with the evolution of the peak frequency during the integration time of each flux states.
To compare the source energetics under these two emission models, we estimate the jet power by assuming the inertia of the
jet is mainly provided by cold protons with their number equal to that of non-thermal electrons. The total jet power due to protons, electrons and the magnetic field will be \citep{Celotti2008}
\begin{align}
P_{\rm jet}= \pi R^2\Gamma^2\beta_\Gamma c(U_e+U_p+U_B)
\end{align}
In Table \ref{table:sed}, we provide $P_{\rm jet}$ for different flux states under EC/BLR and EC/IR $\gamma$-ray emission models along with the 
total radiated power $P_{\rm rad}$. Low value of $\Gamma$ obtained in the case of EC/BLR interpretation, results in very low 
$P_{\rm jet}$ which is smaller than $P_{\rm rad}$. This indicates the jet lose all its power at the initial stage itself 
and questions the existence of large scale AGN jets. Based on these, results we conclude the $\gamma$-ray emission mechanism is
associated with the EC/IR process and plausibly the enhancement in the bulk Lorentz factor being the main cause of the observed 
multi-wavelength flare.

 On comparing our results with the previous flaring studies of 3C 279, we noted that the $\gamma$-ray emission region to be located outside BLR region is consistent with most of the studies (see e.g \citealp{Sahayanathan2012, Dermer2014, Yan2015, Vittoini2017}). However, the exceptional properties  of 3C\,279 like 2015 June flaring (see Introduction) demands a very compact emission region with high density of photons, such regions are mostly located at the BLR region \citep{Hayashida2015, Ackermann2016, Pittori2018}. Even though we found that the observed SEDs of previous flaring events are explained satisfactorily by various emission models, the number of parameters in these models (including in our model) exceeds the number of observables. This makes it hard to obtain similar set of parameters even for the same flaring event with different models. However, the comparative study between the best-fit model parameters of various flaring epochs will help us to identify the dominant physical parameters responsible for the flare emission. Therefore, we compare our best-fit model parameters (see Table \ref{table:sed}) with the parameters obtained in 2013 December, 2014 April and 2015 June flares. The 2013 December flare was studied in detail by \citep{Paliya2015c} using the two-zone leptonic model. Between the low and high flux states (flare 2 and flare 1), the derived model parameters show an increasing trend, in particular $\Gamma$ from 30 to 45 and B from 0.12 to 0.20 G. We also found that the enhancement in the flux from QS to FS is due to increase in $\Gamma$ (from 24.02 to 38.44), however, the obtained values of B from our model showed a marginal decreasing trend (from 0.60 to 0.52 G).
\citet{Yan2015} separately modelled the 2013 December flaring by using the one-zone model with log parabola particle distribution and suggested that the emission region is located within IR region. They also found an increase in $\Gamma$ (18 to 28) from low to high flux state, while magnetic field in both the states is nearly equal to $\sim 1G$. Further, they reported that the radiative power is large compared to magnetic and particle power, which is consistent with our results. Using the same model as ours, \citet{Sahayanathan2018} studied the 2014 March-April flaring epoch ($\sim 3$ times less bright than 2018 January flaring) and derived  smaller $\Gamma$ ($\sim 25.45$) and B ($\sim 0.41$) compared to ours. Recently, \citet{Pittori2018} used an one-zone model with double power law particle distribution to model the 2015 June flare. This study suggested the emission region at location of BLR and lower values of $\Gamma\sim20$. However, \cite{Ackermann2016} reported a $\Gamma\sim50$ and very low magnetization for the standard external radiation Comptonization scenario, in order to satisfy the the minute timescale variability and to avoid overproducing of SSC component. These studies together with our results, suggest that $\Gamma$ plays an important role in the flaring and always shows an increasing trend from low to high flux state irrespective of the model. 


\begin{table}
\centering
\large
\begin{tabular}{@{}lccccccr}
\hline 
& \multicolumn{3}{c}{IR photons} && \multicolumn{3}{c}{BLR photons} \\ \cline{2-4} \cline{6-8}
Parameters & FS & PS & QS &&  FS & PS & QS \\
\hline 
\vspace{1.5mm}
p & $1.18_{1.10}^{1.41}$ & $1.70_{1.18}^{1.77}$ & $1.56_{1.12}^{2.00}$ &&  1.38 & 2.04 & $2.26_{1.42}^{2.71}$\\
\vspace{1.5mm}
q & $4.21_{4.03}^{4.43}$ &  $4.16_{3.95}^{4.33}$ & $4.30_{4.04}^{4.51}$ && 3.93 & 4.12 & $4.54_{4.10}^{5.24}$\\
\vspace{1.5mm}
$\Gamma$ & $38.44_{22.00}^{43.95}$ & $ 30.16_{19.30}^{41.00}$ & $24.02_{15.90}^{26.01}$ && 2.57 & 3.02 & $3.75_{2.50}^{4.90}$\\
\vspace{1.5mm}
$B$ & $0.52_{0.49}^{0.59}$ & $0.73_{0.61}^{0.79}$ & $0.60_{0.51}^{0.67}$ && 8.30 & 13.69 & $11.58_{8.69}^{14.50}$\\
\vspace{1.5mm}
$\rm U_e$ & $0.40_{0.33}^{0.47}$ & $0.20_{0.18}^{0.22}$ & $0.15_{0.13}^{0.16}$ && 1.16 & 0.40 & $0.22_{0.13}^{0.31}$\\
\hline
Properties &&& \\ 
\hline
\vspace{1.5mm}
$P_{\rm jet}$ & 46.42 & 46.10 & 45.74 && 44.77 & 44.98 & 45.01\\
\vspace{1.5mm}
$P_{\rm rad}$ & 43.53 & 43.11 & 43.14  && 46.72 & 46.11 & 45.85\\
\vspace{1.5mm}
$U_e/U_B$ & 37.09 & 9.32 & 10.66 && 0.43 & 0.05 & 0.04\\
\hline
\end{tabular}
\caption{Best-fit source parameters and properties of 3C\,379 obtained from FS, PS, and QS by considering target photon field from IR region (with temperature 1000 K) and BLR region (with frequency $\rm 2.47\times10^{15} Hz$), using the local XSPEC emission model developed by \citet{Sahayanathan2018}. Row:- 1, 2: low energy and high energy power-law index of the particle distribution, 3: bulk Lorentz factor of the emission region, 4: magnetic field in units of $\rm G$, 5: particle energy density in units of $\rm erg\,cm^{-3}$, 6: logarithmic jet kinetic power derived from the source parameters assuming equal number of cold protons as of non-thermal electrons in units of $\rm{erg\,s^{-1}}$, 7: logarithmic total radiated power derived from the source parameters in units of $\rm{erg\,s^{-1}}$ and 8: ratio of particle energy density and magnetic field energy density. Following parameters were kept fixed in all the three states (FS, PS and QS): size of emission region $R'$ (cm), target photon temperature $\rm T$ in K, minimum and maximum Lorentz factor of particle distribution $\rm \gamma_{min}$ and $\rm \gamma_{max}$ respectively, electron Lorentz factor corresponding to the break energy $\rm \gamma_b$, viewing angle $\theta$, covering factor $\rm f$. In case of IR region, $R=10^{16}$, $\rm T=1000\, K$, $\rm \gamma_{min}=60$, $\rm \gamma_{max}=10^6$, $\rm \gamma_b=1500$, $\theta=2^\circ$ and $\rm f=0.10$. While as for target photon field from BLR region, $R=10^{16}$, $\rm T=42000\, K$, $\rm \gamma_{min}=60$, $\rm \gamma_{max}=10^6$, $\rm \gamma_b=1500$, $\theta=7^\circ$ and $\rm f=0.001$. The subscript or superscript values on parameter are upper and lower values of model parameters obtained in spectral fitting.}
\label{table:sed}
\end{table}

\section{Summary}

We studied the broadband temporal and spectral properties of 3C\,279 during its flaring state in 2018 January. The simultaneous \emph{Fermi}-LAT and {\it Swift} observations  allowed us to build a  multi-wavelength light curve and simultaneous broadband SEDs. The temporal analysis of MLC suggests that the source exhibits significant variability at all energy bands, with variations larger at $\gamma$-ray energy than the X-ray and optical/UV bands.  The temporal analysis of $\gamma$-ray emission shows a delay of $\sim 1.0$\,d between the peaks of low energy (0.1--3 GeV) and high energy (3--500 GeV) light curves. A similar timescale lag was observed in the 2015 June flaring between the degree of optical polarization and the optical flux \citep{Pittori2018}. These results suggest that the lag at $\gamma$-ray energies is possibly related to the behaviour of orientation of magnetic field in the jets. In addition, these light curves show asymmetry with a slow rise--fast decay trend in the 0.1--3 GeV energy range and fast rise--slow decay trend in the 3--500 GeV band. The asymmetry in the $\gamma$-ray light curve can not be explained alone with strengthening and weakening of the underlying acceleration mechanism which would lead to similar behavior in both the light curves. A plausible reason may be associated with the shift in SED peak energy during the flare, which is supported by the negative correlation observed between $\alpha$ and $\beta$; $\alpha$ and flux, and positive correlation between flux and energy corresponding to the Compton SED peak.
 During the VHE detection epoch, the emission from 3C\,279 is associated with hard $\gamma$-ray photon index and large curvature values, which indicate that shift in high energy SED peak towards larger energy results in enhancement of the VHE flux.
Further, the spectral properties and the source behavior are investigated by choosing three flux states from the MLC viz. FS , PS and QS where simultaneous observations in $\gamma$-ray, X-ray and optical/UV energies are available. The broadband SEDs of the three flux states are modelled under synchrotron, SSC and EC emission processes and by employing $\chi^2$-minimization technique. 
We considered two possible cases of seed photons for the EC process, namely the thermal IR emission from the dusty torus 
and the monochromatic emission from the BLR region. The broadband SED during different flux states can be fitted reasonably 
well in both cases; however, the parameters obtained when the $\gamma$-ray spectra is considered due to EC/IR mechanism 
are more acceptable than EC/BLR mechanism. 
The later detection of VHE emission from the source further endorse the EC/IR mechanism as the plausible gamma-ray emission
emission mechanism and advocates the emission region to be beyond the BLR region. 
Based on these results, we conclude that the gamma ray emission in 3C\,279 during January 2018 flaring is due to inverse 
Compton scattering of the IR photon from the dusty torus. Further, the flux enhancement is mainly due to the increase in bulk 
Lorentz factor of the jet which is supported by the spectral modelling and the shift in SED peak inferred from the 0.1--3 GeV
light curve.  The comparison of our results with the previous studies suggests that the increasing trend in the bulk Lorentz factor has been observed in all the flaring events irrespective of the model. This further indicates that the bulk Lorentz factor plays a vital role in the flaring activity.

\section*{Acknowledgements}
 We thank the anonymous referee for the constructive comments and suggestions that significantly improved this manuscript. ZS and VJ thank Jianeng Zhou for the useful discussion. We acknowledge the use of data from {\it Fermi} Science Support Center (FSSC) and {\it Swift} data from the High Energy Astrophysics Science Archive Research Center (HEASARC), at NASA’s Goddard Space Flight Center. This research has made use of the XRT Data Analysis Software (XRTDAS) developed under the responsibility of the ASI Science Data Center (ASDC), Italy. ZS acknowledges the support from IUCAA under the visitor's program as most part the research work is carried with this program.

\end{document}